\documentclass[prl,amssymb,
twocolumn,
showpacs]{revtex4}

\usepackage{graphicx}
\usepackage{bm}

\begin{document}

\title{Fidelity and quantum chaos in the  mesoscopic device for
the Josephson flux qubit}

\author{Ezequiel N. Pozzo}
\author{Daniel Dom\'{\i}nguez}
\affiliation{Centro At{\'{o}}mico Bariloche and Instituto Balseiro,
8400 San Carlos de Bariloche,
R\'{\i}o Negro, Argentina.}

\begin{abstract}

We show that the three-junction SQUID device designed for the
Josephson flux qubit can  be used to study the dynamics of quantum
chaos when operated at high energies. We  determine the parameter
region where the system is classically chaotic.
We  calculate numerically the fidelity or
Loschmidt echo (LE) in the quantum dynamics under perturbations in the
magnetic field and in the critical currents, and study different regimes of the LE.
We discuss how the LE could be observed experimentally considering both
the preparation of the initial state and the measurement
procedure.

\end{abstract}


\maketitle

Ultrasmall Josephson devices  have been used as tools for studying
quantum phenomena at the
macroscopic level since the 1980s \cite{legget}. Macroscopic
quantum tunneling \cite{mqt} and macroscopic quantum coherence of
flux \cite{friedman} and persistent currents \cite{vanderwal} have
been observed experimentally. More recently, mesoscopic Josephson
devices have been used for the design of qubits
for quantum computation
\cite{revqubits,qbit_mooij,chiorescu,fqubit_recent}. The
progress made in this case allows to have nowadays Josephson
circuits with small dissipation and large decoherence times,
giving place to a coherent manipulation of the system
\cite{chiorescu,fqubit_recent}.
This
could also make possible the use of Josephson devices for the
study of the quantum dynamics of 
chaotic systems, a subject where
 there has been
a great interest
in the last years
\cite{peres,jp,jacquod,levarios,cucchietti03,jacquod05}.

The stability of the quantum dynamics
of a system against perturbations
\cite{peres} can be quantified
by the fidelity or Loschmidt echo (LE) \cite{jp}.
The LE is the overlap between two states that evolve from the
same initial wave function $|\Psi_0\rangle$ under two slightly
different hamiltonians,
\begin{equation}
F(t)=|f(t)|^2
= |\langle \Psi_0|e^{i H_\varepsilon t/\hbar}
e^{-iH_0t/\hbar}|\Psi_0\rangle|^2,
\end{equation}
with $H_0$, $H_\varepsilon$ the unperturbed and perturbed
hamiltonians, respectively.
In classically chaotic systems,
above a perturbative regime where the LE has
a Gaussian decay for short times,
 the LE shows an exponential decay for
large $t$, $F(t)\propto e^{-\gamma t}$ \cite{jp,jacquod,levarios,cucchietti03,jacquod05}.
For  weak perturbation strength $\varepsilon$
the decay constant $\gamma$  depends as
$\gamma(\varepsilon) \propto \varepsilon^2$,
and is obtained from
the Fermi golden rule (the FGR regime) \cite{jacquod},
while for strong perturbations
$\gamma$ becomes independent of $\varepsilon$ and
saturates at the classical Lyapunov exponent $\lambda$ (the Lyapunov regime) \cite{jp}.
There  have been some recent experimental measurements of the LE
\cite{andersen,schafer,ryan}.
Here we will show that
the device for the Josephson flux qubit (DJFQ) studied
in \cite{qbit_mooij,chiorescu,fqubit_recent}
is also a promising system for the experimental observation
of the LE.

The DJFQ consists of three
Josephson junctions in a superconducting ring \cite{qbit_mooij}
that encloses a magnetic flux $\Phi= f\Phi_0$,
with $\Phi_0=h/2e$.
Two of the junctions have the same coupling
energy $E_J$ and capacitance $C$, while the
third junction has couplings $\alpha E_J$ and $\alpha C$, respectively
($0.5<\alpha<1$).
Typically the circuit inductance
can be neglected  and
the phase difference of the
third junction is:
$\varphi_3=2\pi f+\varphi _1 -\varphi _2$,
leading to
the Hamiltonian \cite{qbit_mooij}
\begin{equation}
{\cal H}=\frac{1}{2}{\vec {P}}^T
{\rm {\bf M}}^{-1}{\vec {P}}
+E_J V(\vec {\bf \varphi})\label{ham_clas}
\end{equation}
where $\vec{\varphi}=(\varphi_1,\varphi_2)$,
$\vec{P} = {\rm{\bf M}}\cdot d{\vec{\varphi}}/dt$,
and
$$
{\rm {\bf M}}= {\left(\frac{\Phi_0}{2\pi}\right)^2}C\left(
{{\begin{array}{cc}
 {1+\alpha +\gamma }  & {-\alpha }  \\
 {-\alpha }  & {1+\alpha +\gamma }  \\
\end{array} }} \right)={\frac{\hbar ^2}{\eta^2E_J}}{\rm {\bf m}}
.$$
Here $\eta=\sqrt{8E_C/E_J}$ with $E_C=e^2/2C$,
we include in  $\bf M$ the  on-site gate capacitance $C_g=\gamma C$, and
\begin{equation}
V(\vec {\bf \varphi})=
2+\alpha -\cos \varphi_1-\cos \varphi_2
- \alpha \cos (2\pi f+\varphi _1 -\varphi _2 )
\end{equation}
In the quantum
regime,
$\hat{\vec{P}}= -i\hbar\nabla_\varphi = -i\hbar(\frac{\partial}{\partial\varphi_1},\frac{\partial}{\partial\varphi_2})$,
the time-dependent Schr\"odinger equation  is
\begin{equation}
i\eta\frac{\partial \Psi(\vec {\bf \varphi})}{\partial t}
= \left[ -\frac{\eta^2}{2}\nabla_\varphi^T{\rm{\bf m}}^{-1}\nabla_\varphi
+V(\vec {\bf \varphi})\right] \Psi(\vec {\bf \varphi})
\end{equation}
where we normalized time by $t_c
=\hbar/\eta E_J$, 
energy by $E_J$ and momentum by
$\hbar/\eta$. We see in Eq.(4)
that the parameter $\eta$ plays the role of an effective
$\hbar$.
For quantum computation implementations \cite{qbit_mooij,chiorescu,fqubit_recent}
the DJFQ is operated at magnetic
fields near the half-flux quantum ($f= 1/2+\delta f$).
In this case the two lowest energy eigenstates are
symmetric and antisymmetric superpositions of two
states corresponding to macroscopic persistent currents
of opposite sign. These two eigenstates are energetically separated
from the others (for small $\delta f$)
and therefore the DJFQ has been  used as a qubit \cite{qbit_mooij,chiorescu,fqubit_recent}.
As we will discuss here, the higher energy states
of the DJFQ show quantum manifestations of classical chaos.

\begin{figure}[th]
\begin{center}
\includegraphics[width=0.9\linewidth]{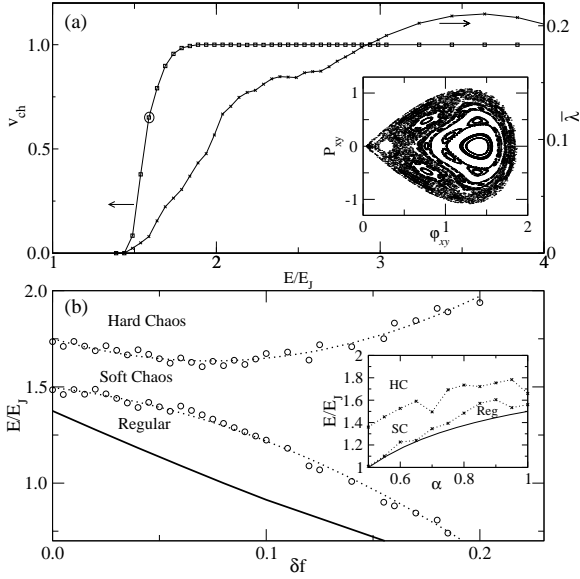}
\caption{ (a) Chaotic volume $v_{ch}$ and average Lyapunov
exponent $\overline{\lambda}$ versus energy $E$ for $\alpha=0.8$
and $f=1/2$. Inset: surface of section
$\varphi_{xy}=(\varphi_1-\varphi_2)/\sqrt{2}$,
$P_{xy}=(P_1-P_2)/\sqrt{2}$ at $E=1.6E_J$. (b) Energy boundaries
for the regimes of regular orbits, soft chaos and hard chaos as a
function of $\delta f=f - 1/2$ for $\alpha=0.8$. The
continuous line corresponds to the potential energy minimum,
$E_{min}$. Inset: Energy boundaries for the different regimes as a
function of $\alpha$ for $f=1/2$.} \label{fig1}
\end{center}\end{figure}
\noindent

It has been found in \cite{parmenter,kato} that superconducting
loops with three Josephson junctions  and on-site capacitances
($\gamma=\infty$) can be chaotic. Most Josephson circuits, like the DJFQ, 
have small on-site gate capacitances ($\gamma \sim 10^{-2}$). 
Here we analyze the dynamics of the DJFQ
considering the realistic case with $\gamma=0.02$
\cite{qbit_mooij}. We first obtain the classical dynamical
evolution integrating the Hamilton equations that correspond to
Eq.(2)  with a second order Verlet algorithm. For different values
of the parameter $\alpha$ and magnetic field $ f $ we compute the
maximum Lyapunov exponent $\lambda$ for each orbit at different
energies $E$. We estimate the chaotic volume $v_{\rm ch}(E)$, as
the probability of having a chaotic orbit (i.e. $\lambda > 0$) for
a given $E$, using $10^3$ initial conditions randomly chosen with
uniform probability within the available phase space. 
Also the average Lyapunov exponent, $\bar{\lambda}(E)$, of
the chaotic orbits is obtained. In Fig.1(a) we show $v_{\rm
ch}(E)$ and $\bar{\lambda}(E)$ for $f=0.5$ and $\alpha=0.8$. Above
the minimum energy of the potential, $E_{min}$, we find: (i) {\it
regular orbits} for $E_{min}<E<E_{ch}$ ($v_{\rm ch}=0$), (ii) {\it
soft chaos} ({\it i.e.}, coexistence of regular and chaotic
orbits, $0<v_{\rm ch}<1$) for $E_{ch}<E<E_{hc}$, a Poincare
section for this case is shown in the inset of Fig.1(a), and (iii)
{\it hard chaos} (all orbits are chaotic, $v_{\rm ch}=1$) for
$E>E_{ch}$. The average Lyapunov exponent is $\bar{\lambda}>0$
above $E_{ch}$.
In Fig \ref{fig1}(b) we show the energy boundaries of the
different regimes ($E_{min},E_{ch},E_{hc}$) as a function of
$\delta f=f-1/2$ for $\alpha=0.8$. We can see that for $f=1/2$ the
onsets of soft and hard chaos, $E_{ch}$, $E_{hc}$, are closer to
$E_{min}$ in comparison with other values of $f$. In the inset of
Fig.\ref{fig1}(b) we also show the boundaries of the dynamic
regimes as a function of $\alpha$ for $f=1/2$. The hamiltonian
dynamics of Eq.(2) are a good approximation of the problem for
energies $E<2\Delta$, with $\Delta$ the superconducting gap.
The Al/AlO$_x$/Al junctions of \cite{qbit_mooij} have $2\Delta\approx3.7E_J$.
Since we find $E_{hc} = 1.75 E_J$, there is a wide energy range where the
hard chaos regime is experimentally accessible. 
 Furthermore,
we find that for realistic experimental parameters ($\eta = 0.1 -
0.5$, $\alpha=0.7-0.8$), the third quantum energy level can be
above the classical onset of chaos, $E_{ch}$, for $f=1/2$. 
In Ref.~\cite{kato}, for the case  with $\gamma=\infty$,
an analysis of the statistics of the energy spectra of
the quantum hamiltonian shows that it belongs to the gaussian orthogonal ensemble
in the hard chaos regime.
We find a similar result in  our case, with $\gamma=0.02$ and $f=1/2$, 
for $E > E_{hc}$ \cite{unp}.

\begin{figure}[tbh]
\begin{center}
\includegraphics[width=0.9\linewidth]{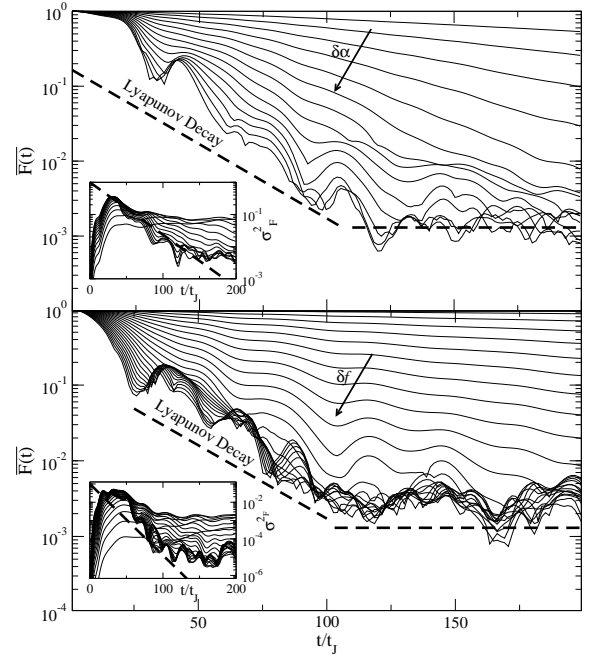}
\caption{
The average Loschmidt echo, $\overline{F}(t)$, as a function of time for
different values of the perturbation strength of (a) $\delta\alpha$  and (b) $\delta f$.
The energy  is $E\simeq3E_J$, deep in
the hard chaos region, and $\hbar_{\rm eff}\equiv\eta=\sqrt{\frac{8E_C}{E_J}}=0.17$.
 Time is measured in units of
$t_J=\hbar/E_J=\eta t_C$. Insets: time evolution of
$\sigma_F^2=\overline{F^2(t)}-\overline{F(t)}^2$. The dashed lines
represent an exponential decay given by the Lyapunov exponent
($\lambda=0.182 \pm 0.008$) with the $t=0$ offset shifted for
clarity.} \label{fig2}
\end{center}\end{figure}
\noindent

We now calculate the quantum dynamics of the DJFQ integrating numerically
Eq.~(4) with  a fourth order split-operator algorithm  as in
\cite{feit}, with a discretization grid of $\Delta\varphi=2\pi/128$ and $\Delta t= 0.015\eta t_C$.
We use $2\pi$-periodic boundary conditions on $\vec {\bf \varphi}=(\varphi_1,\varphi_2)$.
To compute the LE of Eq.(1), the simulations are started from
minimum-uncertainty
$2\pi$-periodical wave packets \cite{carruthers} given by
\begin{equation}
|\Psi_0\rangle=\Psi_{\vec{K_0},\vec{\varphi_{0}}} (\vec{\varphi})
= C\cdot
e^{i \vec{K_0}\cdot \left(\vec{\varphi}-\vec{\varphi_0}\right)}\cdot
e^{-B(\vec{\varphi}-\vec{\varphi_0})/{2\sigma^2}}\label{paquete}
\end{equation}
with  $B(\vec{\varphi})=2-\cos\varphi_1-\cos\varphi_2$,
[$B(\vec{\varphi})\approx |\vec{\varphi}|^2/2$ for small
$|\vec{\varphi}|$], $\vec{K_0}=(k_1,k_2)$ with $k_1,k_2$ integers,
and $\sigma$ is the width of the  wave packet. The LE is usually
computed as the average $\overline{F}(t)$ over different 
$|\Psi_0\rangle$ (see \cite{levarios,jacquod05}). Here we average
over 15 initial conditions with different
$\vec{K_0},\vec{\varphi_{0}}$ corresponding to the same classical
energy.
We choose $\sigma^2=0.31\eta$, which
 corresponds to a spectral width of $\Delta E\approx 0.3 E_J$.
We consider $E = 3 E_J > E_{hc}$, for which the classical phase
space is filled by a  connected region of chaos with Lyapunov
exponent $\lambda = 0.182 \pm 0.008$.
 

\begin{figure}[tbh]
\begin{center}
\includegraphics[width=0.9\linewidth]{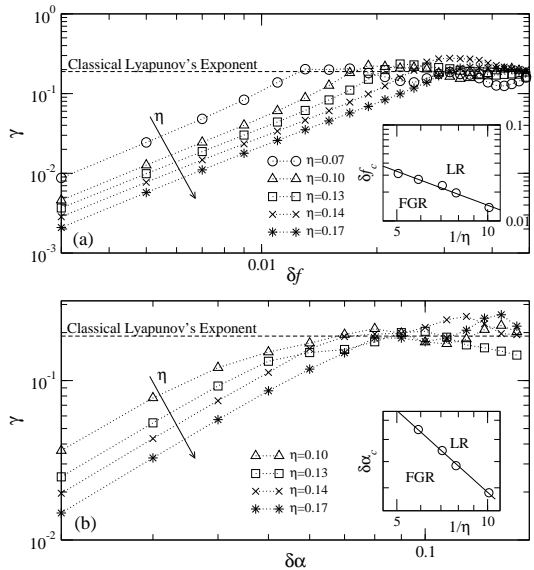}
\caption{(a) Decay rate of the LE, $\gamma$,
 as a function of the perturbation strength $\delta f$
for different values of $\eta$, with $E\simeq3E_J$.
The classical  Lyapunov exponent is indicated in the dashed line.
Inset: crossover perturbation  $\delta f_c$ as a function of $\eta^{-1}$.
(FGR: Fermi golden rule regime, LR: Lyapunov regime).
 (b)  Same as (a) for perturbations $\delta\alpha$.} \label{fig3}
\end{center}\end{figure}
\noindent

We evaluate  the LE of the
DJFQ against perturbations in the parameters $\alpha$  and $f$.
In Fig.\ref{fig2}(a) we show the time dependence of
$\overline{F}(t)$ for different perturbations in the parameter
$\alpha=\alpha_0 + \delta\alpha$ for $\alpha_0=0.8$, while in Fig.
\ref{fig2}(b) we show $\overline{F}(t)$ for different
perturbations in $f=1/2+ \delta f$. We can see that
in both cases $\overline{F}(t)$ decays with time, and that the
decay rate tends to  increase when increasing the perturbation.
Above a crossover value of the perturbation
($\delta\alpha_c$  or $\delta f_c$) we find that the curves of
$\overline{F}(t)$ tend to overlap.  The overall decay of the LE is
has been described with the form $\overline{F}(t)\sim Ae^{-\gamma
t} + F_\infty$ \cite{levarios}. For large perturbations the decay
rate $\gamma$ saturates at a value close to the Lyapunov exponent
$\lambda$ of the classical dynamics. This can be seen in Fig.2
where the dashed lines show the slope of a decay rate with the
Lyapunov exponent for comparison. The constant value $F_\infty$ is
proportional to the inverse of the fraction of the volume of the
Hilbert space spanned by the initial wave function $|\Psi_0\rangle$ \cite{levarios}.
In our case this corresponds to
$F_\infty\propto \sigma^2/(2\pi)^2 = 0.31 \eta/(2\pi)^2\approx 0.0013$
for $\eta=0.17$, which is 
close to the results of Fig.~\ref{fig2}(a),(b).
An analysis of the variance of
the fidelity, $\sigma_F^2(t) =
\overline{F(t)^2}-\overline{F(t)}^2$ is also important \cite{jacquod05}.
This is shown in the insets
of Fig.~\ref{fig2}.
We find that for increasing perturbations  $\sigma_F^2$
saturates to a
decay given by $\sigma_F^2(t) \sim e^{-2\lambda t}$,
as discussed in \cite{jacquod05}.

We obtain the decay rate $\gamma$ fitting the exponential form
$Ae^{-\gamma t} + F_\infty$ for $\overline{F}(t)$ for times above
the initial gaussian decay (we have chosen $t > 15 t_J$ in this
case). Fig. \ref{fig3}(a) shows the obtained $\gamma$ as a
function of the perturbation $\delta f$ for different values of
$\eta$. For small perturbations we obtain a quadratic law
dependence of the decay rate with the perturbation strength,
$\gamma \propto (\delta f)^2$, which corresponds to the Fermi
Golden Rule (FGR) regime \cite{jacquod}. For large perturbations
the  obtained values of $\gamma$ 
have a large error, 
which is of the size of
the oscillations seen in the data in Fig.\ref{fig3}(a) 
for large $\delta f$.  
However, 
when comparing different cases of $\eta$, 
we see that for large perturbations the values of
$\gamma$ fall close to the 
Lyapunov exponent (shown as a dashed line).
We obtain an estimate of the crossover value $\delta f_c$ where
$\gamma$ saturates to $\lambda$. In the inset of Fig.
\ref{fig3}(a) we see the dependence of $\delta f_c$ with $\eta$.
We find that $\delta f_c$ decreases for decreasing $\eta$ since
quantum fluctuations become less important, and therefore the
classical Lyapunov decay is reached more easily.
We find a similar behavior for the dependence of $\gamma$
with  $\delta\alpha$ and $\eta$ [shown in Fig.\ref{fig3}(b)].

\begin{figure}[tbh]
\begin{center}
\includegraphics[width=0.9\linewidth]{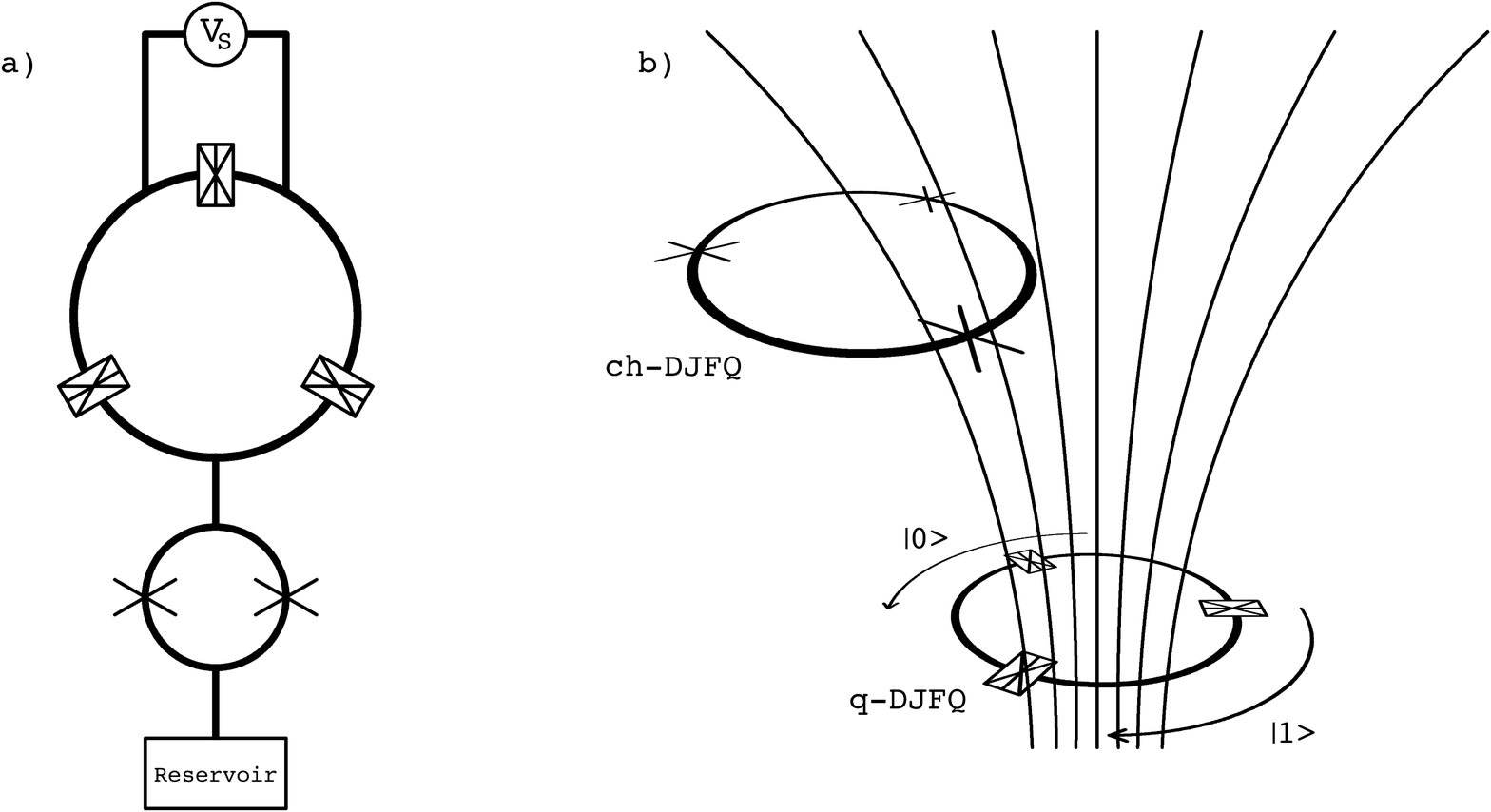}
\caption{Schematic setup for the observation of the
Loschmidt echo in a
DJFQ.
(a) Circuit for preparation of the initial wave packet. $V_s$ is
a voltage source
and the SQUID connects a node of the DJFQ with a superconducting
reservoir.
Only  one voltage source and one SQUID are drawn, for
simplicity.
(b) Layout for the measurement of the Loschmidt echo.
ch-DJFQ: circuit
with  quantum evolution in the chaotic regime;
q-DJFQ: circuit used as a qubit for measurement of the LE.} \label{fig4}
\end{center}\end{figure}
\noindent

Therefore, we have shown that the DJFQ at high energies shows a
decay of the LE,  similar to other chaotic systems
\cite{jp,jacquod,levarios,cucchietti03,jacquod05}. This behavior
could be experimentally observable if the time scale for the
Lyapunov decay, $\tau_{\rm Lyap}$ is much smaller than the
decoherence time $\tau_{\rm decoh}$ due to external sources.  We
estimate $\tau_{\rm Lyap}=\hbar/(\eta E_J \lambda)\sim 0.04 -
0.2$ns, using $\lambda\approx 0.18$ from the hard chaos regime and
experimental parameters \cite{chiorescu,fqubit_recent}. In
\cite{chiorescu} a value of $\tau_{\rm decoh}\approx 20$ns was
obtained for the lowest  energy states.
At high energies ($E \sim 3 E_J$),
we find that  the spectrum is one order of magnitude more dense than at low energies,
therefore $\tau_{\rm decoh}$ should be at least one order of
magnitude smaller,  $\tau_{\rm decoh} \sim 1$ns. These roughly
estimated  values of $\tau_{\rm Lyap}\sim0.1$ns and $\tau_{\rm
decoh}\sim1$ns leave some room for observing the FGR and Lyapunov
regimes of the LE. However, in order to realize an experiment, two
issues have to be solved: (i) preparation of the initial state and
(ii) measurement of the fidelity $F(t)$. (i) {\it Preparation of
the initial state}.
In order to observe the Lyapunov regime the system has to be started
from a wave packet narrowly localized in both coordinate (phase)
and momentum (charge) \cite{jp,jacquod}, as in Eq.(5). Here we suggest the
procedure shown schematically in Fig.4(a). To localize the momentum
(charge): each of the junctions is connected in parallel to a
voltage source $V_s$, which builds up a charge in each junction.
(For $E=3E_J$ a
voltage of $V_s
\approx 0.1$ mV is
needed). To localize the coordinate (phase):  each of the nodes of the DJFQ is
connected to a large superconducting reservoir through a DC-SQUID
as in \cite{elion}.
Then the system is prepared staying with $V_s\not=0$ and
$\Phi_{\rm SQUID}=0$ for a long time ($t<0$),
and at $t=0$ the voltage sources are set to $V_s=0$ and the
flux in the SQUIDs to $\Phi_{\rm SQUID}=\Phi_0/2$.
(ii) {\it Measurement of the fidelity}.
One has to be able to measure the overlap of Eq.(1).
One protocol proposed originally in \cite{gardiner} and
applied in \cite{montangero05,lesovik} (see also
\cite{andersen}) consists in coupling the chaotic system under
consideration with a qubit that acts both as a perturbing and a
measuring device. The approximate hamiltonian for this case is
$H=H_0 \otimes|0\rangle\langle 0| + H_\epsilon \otimes|1\rangle\langle1|$.
$H_0$, $H_\epsilon$ are the hamiltonians of the unperturbed and perturbed  system,
respectively, and $|0\rangle$, $|1\rangle$ are the two basis states of the
qubit such that  when the qubit is in the $|1\rangle$ state it induces a
perturbation $\epsilon$ in the chaotic system. A Ramsey type of
experiment is performed:
after a time $t$,  $\pi/2$  pulses are  applied to the qubit
for two evolutions from initial states of the system
$|\Psi_0\rangle\otimes(|0\rangle+a|1\rangle)/\sqrt{2}$ with $a=1$ and $a=i$ respectively,
from which the fidelity amplitude $f(t)$ can be obtained, see
\cite{gardiner,montangero05,lesovik} for details. A possible
implementation of this idea is shown in Fig.4(b). A second DJFQ
could be used operating as a qubit (called q-DJFQ) for measurement
of the LE in a DJFQ evolving in the quantum chaotic regime (called
ch-DJFQ). The q-DJFQ is coupled inductively to the ch-DJFQ, such
that when the q-DJFQ is in the $|0\rangle$ ($|1\rangle$) state a
clockwise (counter-clockwise) current flowing in it  induces a
positive (negative) perturbation in the magnetic flux threaded by
the ch-DJFQ.  The q-DJFQ should have a smaller area  than the
ch-DJFQ (such that the flux induced in the q-DJFQ by the currents
in the ch-DJFQ is small). For a better observation of the LE the
ch-DJFQ should be  built in a more semiclassical regime ($\eta
\sim 0.1$, for example), while the q-DJFQ is built in a more
quantum  regime ($\eta \sim 0.5$, for example). Furthermore,
making measurements with the measuring q-DJFQ placed at different
distances from the ch-DJFQ could allow to obtain the LE for
different perturbation intensities. For example, if when placing
the q-DJFQ closer to the ch-DJFQ the decay of the LE becomes
independent of the distance, that would indicate that the Lyapunov
regime was reached. The state preparation and measurement
procedure described here requires coupling to several external
objects, and thus the resulting decoherence rates will inevitably
increase. However, it has been shown in general grounds that $\tau_{\rm
Lyap} \le \tau_{\rm decoh}$ \cite{cucchietti03}. This
implies that at least the Lyapunov regime of the LE will  be
observable, and for this reason it is already interesting to
perform the experiment.
Moreover, one advantage of  Josephson nanocircuits is that they can be fabricated
with well-controlled parameters allowing to study the LE for
different cases of the effective $\hbar$
($\hbar_{\rm eff}\equiv\eta$).

We acknowledge discussions with H. Pastawski, M.J. S\'anchez, H.
Pastoriza and F. Cucchietti and financial support from ANPCYT,
CNEA and Conicet.

\end{document}